\documentclass{PoS}

\usepackage{amsmath}
\usepackage{amssymb}
\usepackage{times}
\usepackage{txfonts}
\DeclareMathAlphabet{\mathbold}{OML}{txr}{b}{it}

\usepackage{wrapfig}

\usepackage{xspace}
\usepackage{multirow}
\usepackage{units}
\usepackage{epsfig}
\usepackage{hhline}
\usepackage{paralist} 

\usepackage{setspace}

\newcommand{\muf}{\ensuremath{\mu_{F}}\xspace}
\newcommand{\mur}{\ensuremath{\mu_{R}}\xspace}
\newcommand{\as}{\ensuremath{\alpha_s}\xspace}
\newcommand{\asmz}{\ensuremath{\alpha_s(m_Z)}\xspace}
\newcommand{\asmur}{\ensuremath{\alpha_s(\mur)}\xspace}

\newcommand{\NNLOJET}{NNLO\protect\scalebox{0.8}{JET}\xspace}

\newcommand{\chisq}{\ensuremath{\chi^{2}}}

\newcommand{\ndf}{\ensuremath{n_{\rm dof}}}

\newcommand{\Qsq}{\ensuremath{Q^2}\xspace}

\newcommand{\ptjet}{\ensuremath{P_{\rm T}^{\rm jet}}\xspace}
\newcommand{\meanpt}{\ensuremath{\langle P_{\rm T} \rangle}}

\newcommand{\pt}{\ensuremath{P_{\rm T}}\xspace}

\newcommand{\etalab}{\ensuremath{\eta^{\mathrm{jet}}_{\mathrm{lab}}}\xspace}

\newcommand{\GeV}{\ensuremath{\mathrm{GeV}}\xspace}
\newcommand{\GeVsq}{\ensuremath{\mathrm{GeV}^2}\xspace}

\title{Determination of the strong coupling at NNLO from jet
  production in DIS}
\ShortTitle{Strong coupling at NNLO from H1 jet cross sections}

\author{\speaker{Daniel Britzger} on behalf of the
  collaboration\thanks{Work performed by the H1
    Collaboration together with V. Bertone, J. Currie, C. Gwenlan,
    T. Gehrmann, A. Huss, J. Niehues and M. Sutton}\\
  DESY, Notkestr. 85, 22607 Hamburg, Germany\\
  E-mail: \email{daniel.britzger@desy.de}}

\abstract{
A first determination of the strong coupling \asmz\ in
next-to-next-to-leading order (NNLO) from inclusive jet and dijet
production cross sections in deep-inelastic scattering at HERA is presented.
Data collected by the H1 experiment in the years 1995 to 2007 covering
the range of momentum transfer $5.5<Q^2<15\,000\,{\rm GeV}^2$ and jet
transverse momenta $\ptjet>4.5\,{\rm GeV}$ are explored.
The strong coupling is determined in a fit to inclusive jet and dijet
data to
$\asmz=0.1157\,(6)_{\rm exp}\,(^{+31}_{-26})_{\rm theo}$.
Further studies on the phenomenological application of the new NNLO
calculations and on fits to the individual data sets are presented.
The running of the strong coupling is probed in a single experiment over one order of
magnitude in the remormalisation scale and consistency with the
QCD expectations is found.
}

\FullConference{XXV International Workshop on Deep-Inelastic Scattering and Related Topics\\
		3-7 April 2017\\
		University of Birmingham, UK}

\begin{document}

\section{Introduction}
The strong coupling constant is one of the least known parameters of
the Standard Model (SM) and a precise knowledge is of crucial importance
for precision physics and searches for physics beyond the SM at the
LHC.
Cross sections for jet production in deep-inelastic electron-proton
scattering (DIS) are directly sensitive to the strong coupling
constant \asmz\ already in leading order in perturbative QCD (pQCD) as
these measurements are performed in the Breit frame of reference.
The cross section calculations are performed in next-to-next-to-leading order (NNLO)
accuracy, where the cross section predictions are obtained with the
program \NNLOJET\,\,\cite{Currie:2016ytq,Currie:2017tpe}.

\section{Methodology}
Cross sections for jet production in $ep$ collisions have been measured by the H1
experiment at HERA at different center-of-mass energies and for
different kinematic regions.
Here, jet and dijet cross
sections taken during the years 1995
to~2007~\cite{Adloff:2000tq,Aaron:2010ac,Aktas:2007aa,Andreev:2014wwa,Andreev:2016tgi}
are considered.
Consistent to all data sets, jets are defined using the $k_t$
jet-algorithm with a parameter of $R=1$, 
and jets are required to be contained in the pseudorapidity range
$-1<\etalab<2.5$ defined in the laboratory frame.
For the selected data, inclusive jet cross sections have been measured double-differentially
as a function of the photon virtuality \Qsq\ and jet transverse
momentum \ptjet, and dijet cross sections as a function of \Qsq\ and
the average transverse momentum of the two hardest jets, \meanpt.
A brief summary of the employed measurements and the kinematic range
of the observables is given in table~\ref{tab:datasets}.
\begin{table}[tbhp]
  \footnotesize
  \begin{center}
    \begin{tabular}{cccccc}
      \hline
      \multicolumn{1}{c}{Data set} & $\sqrt{s}$ & int. $\mathcal{L}$ & DIS kinematic &  Inclusive jets &  Dijets   \\
      \multicolumn{1}{c}{[Ref.]}  & $[\GeV]$   & $[{\rm pb}^{-1}]$  &  range        &                 &   $n_{\rm jets}\ge2 $  \\
      \hline
      $300\,\GeV$\,\cite{Adloff:2000tq} & 300 & 33& $150<\Qsq<5000\,\GeVsq$  &   $7<\ptjet<50\,\GeV$ & $8.5<\meanpt<35\,\GeV$ \\
      HERA-I\,\cite{Aaron:2010ac}    & 319  &43.5 & $5<\Qsq<100\,\GeVsq$   &   $5<\ptjet<80\,\GeV$ & $7<\meanpt<80\,\GeV$ \\
      HERA-I\,\cite{Aktas:2007aa}    & 319  &65.4 & $150<\Qsq<15\,000\,\GeVsq$   &   $5<\ptjet<50\,\GeV$ & $-$  \\
      HERA-II\,\cite{Andreev:2016tgi}& 319  & 290& $5.5<\Qsq<80\,\GeVsq$        & $4.5<\ptjet<50\,\GeV$ & $5<\meanpt<50\,\GeV$  \\
      HERA-II\,\cite{Andreev:2016tgi,Andreev:2014wwa}   & 319  & 351& $150<\Qsq<15\,000\,\GeVsq$     &   $5<\ptjet<50\,\GeV$ & $7<\meanpt<50\,\GeV$ \\
      \hline
    \end{tabular}
    \caption{Summary of the kinematic ranges of the inclusive jet and dijet data taken by the H1 experiment.}
    \label{tab:datasets}
    \end{center}
\end{table}
The data sets are separated into different data taking periods
and two \Qsq-regions, where the scattered lepton is
identified in different experimental devices.
In case of the dijet cross sections, regions of the phase space
exhibiting an infrared sensitivity due to `back-to-back' topologies
are avoided by imposing asymmetric cuts on the transverse momenta of
the two leading jets.

The predictions are calculated as a convolution
of parton density functions (PDFs) and a partonic cross section.
Both these components exhibit a dependence on \asmz\ and their impact
on the results are assessed below.
The partonic cross section has its \as-dependence explicit as it is
calculated in terms of a perturbative expansion in orders of
$\alpha_s^{(n)}$.
The \as-dependence of the PDFs is given by the factorisation theorem
and where it originates from the QCD splitting kernels and the $\beta$-functions. 
Once a PDF is determined for a given value of \asmz\ it can be
translated to any other value of \asmz\ by an integration step.
This translation defines the \as-dependence of the PDF. 
An equivalent solution to this explicit integration is obtained by 
evaluating the PDFs at a suitable value of \muf, which depends on \asmz, thus taking 
full benefit of the factorisation theorem~\cite{thisprelim}.
The PDF parameterisation is based on the NNPDF3.0
PDF set~\cite{Ball:2014uwa}, which was determined for a value of $\asmz=0.118$.
Multiplicative correction factors are applied in order to account for
non-perturbative hadronisation effects.
The renormalisation and factorisation scales are chosen to be
$\mur^2=\muf^2=\Qsq+\pt^2$, where \pt\ denotes \ptjet\ in case of
inclusive jet and \meanpt\ for dijet cross sections.

The value of the strong coupling constant is determined in a fit of these
NNLO calculations to the H1 jet data, where the
\as-dependencies in the predictions, both in the partonic cross
sections and in the PDF, are taken into account. 
The NNLO coefficients are stored in the fastNLO format~\cite{Britzger:2012bs} in order to allow for a
repeated calculation with different values of \asmz\ and different PDF sets.
The fit \chisq-definition accounts for experimental, hadronisation and PDF
uncertainties. Correlations of systematic uncertainties and
statistical correlations of the data are considered.
The uncertainties on the resulting value of \asmz\ due to experimental and
theoretical sources are estimated.
The PDF and hadronisation uncertainties are obtained by repeating the
fit with these uncertainties excluded in the fit and comparing the resulting
fit uncertainty.
The scale uncertainty is estimated by repeating the fit with scale
factors of 0.5 and 2. 
The `PDFSet' uncertainty is obtained as half of the maximum difference
of the results from fits using alternatively the ABMP, CT14,
HERAPDF2.0, MMHT or NNPDF3.0 PDF set, and the `PDF\as' uncertainty is
estimated as half of the difference of the results obtained from fits
using PDFs which were determined with \asmz-values differing by 0.004.


\begin{figure}[t!b!h]
  \begin{minipage}{0.48\linewidth}
    \begin{center}
      \includegraphics[width=0.99\textwidth]{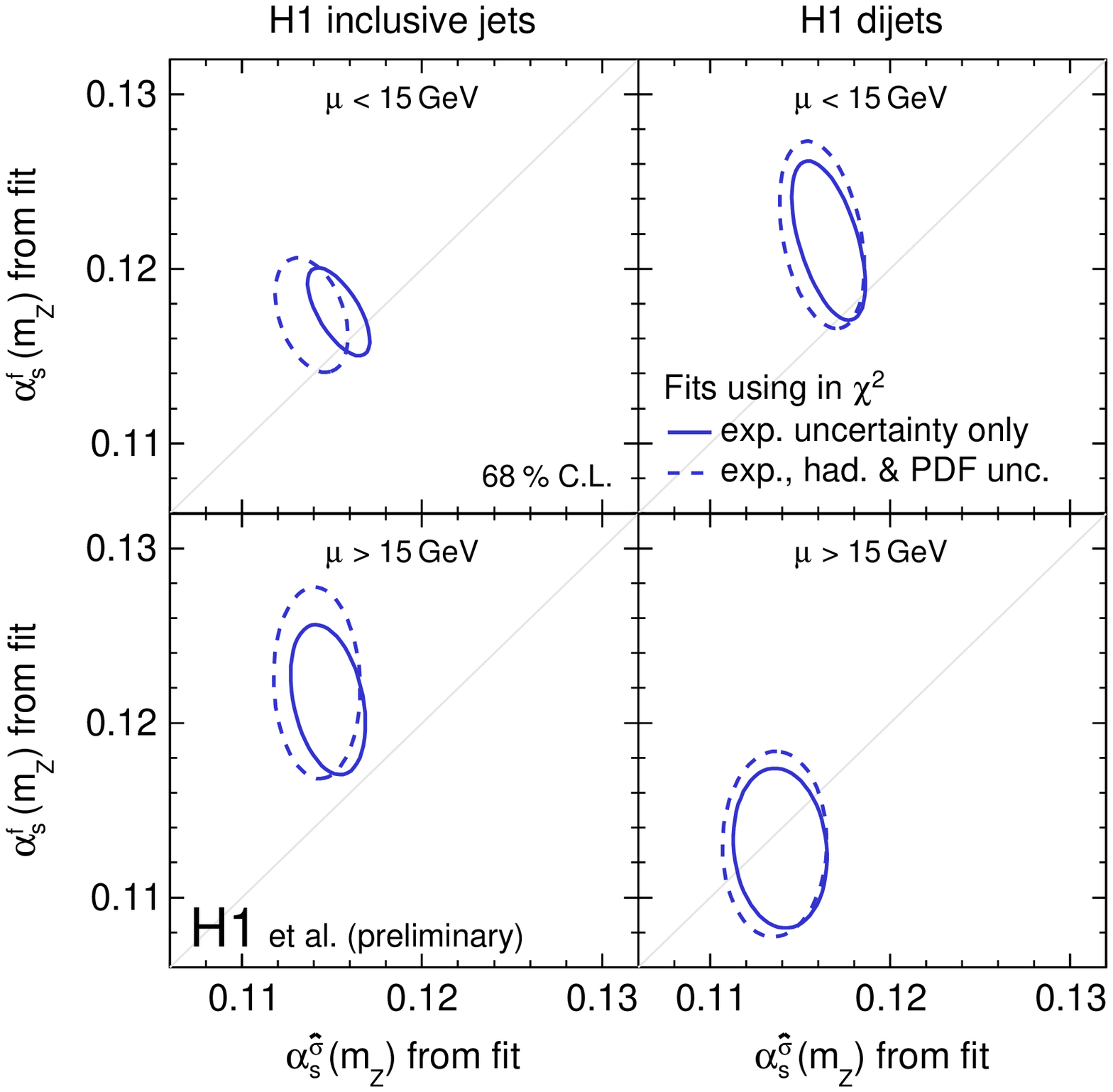}
      \caption{   
      Contours at a confidence level of 68\,\% for fits where the two
      appearances of \asmz\ in the cross section calculation are
      identified separately.
      The upper and lower pads show results from data points with
      \mur\ smaller or greater 15\,\GeV. 
      The dashed contours indicate fits where the PDF uncertainty is not
      considered in the \chisq-calculation.
    }
\label{fig:plot1a}
\end{center}
\end{minipage}
\hskip0.03\linewidth 
\begin{minipage}{0.48\linewidth}
\includegraphics[width=0.99\textwidth]{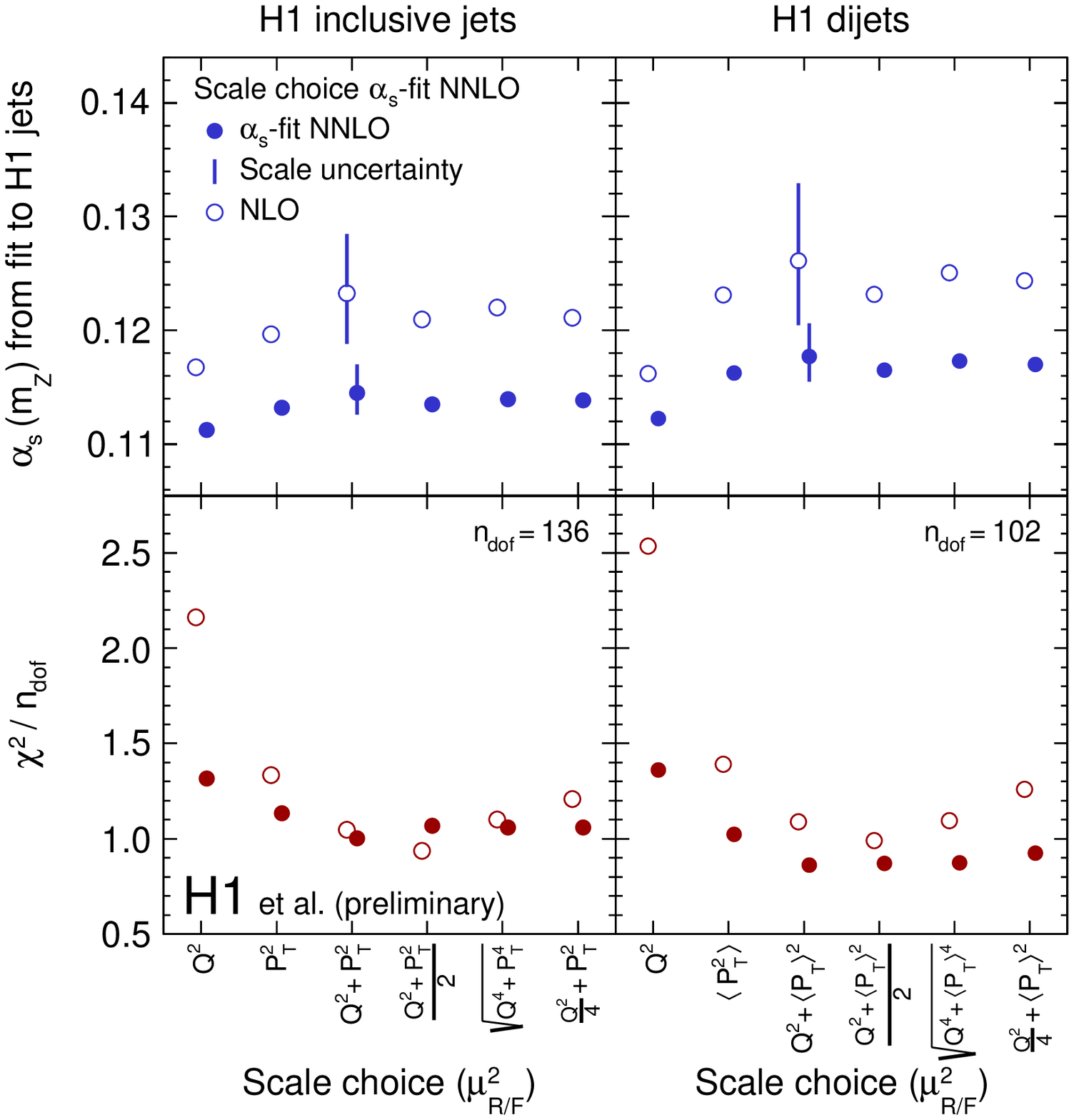}
\caption{   
  Values of \asmz\ obtained from fits to inclusive jet or dijet cross
  sections obtained for different definitions of the renormalisation
  and factorisation scales.
  The lower pads show the values of \chisq/\ndf\ of the fit.
  The open circles display results obtained using NLO matrix elements.
  The vertical error bars indicate the scale uncertainty.
}
\label{fig:plot1b}
\end{minipage}
\end{figure}

\section{Results}
The sensitivity of the data to \asmz\ is studied in fits with two
free parameters representing the 
two \as\ contributions to the calculation, 
assuming those can be chosen independently, i.e.\ one parameter for
the PDFs,  $\as^f(m_Z)$, and another parameter for the hard coefficients, $\as^{\hat\sigma}(m_Z)$.
The fits are performed using inclusive jet or dijet cross section measurements, with data points below or above
the renormalisation value of 15\,\GeV. The contours displaying the
68\,\% confidence level of the fitted results are displayed in figure~\ref{fig:plot1a}. 
The two \asmz\ values determined in the fit are consistent, while
the sensitivity to \asmz\ of the hard coefficients outperforms the one of the PDF.
The two fit parameters are negatively correlated, resulting in an increased
sensitivity for fits using a common \asmz.

Fits are also performed employing alternative definitions for the 
renormalisation and factorisation scales. 
The resulting \as-values and related values of \chisq/\ndf\ for the
individual fits are displayed in figure~\ref{fig:plot1b} for
fits to inclusive jet and to dijet cross sections. 
The results obtained with alternative scale choices are typically
covered by the scale uncertainty. Choosing $\mur^2=\muf^2=\Qsq$ is
disfavored, presumably because this scale is not
sufficiently related to the dynamics of jet production.
For comparison the fits are also repeated with hard coefficients calculated in
NLO accuracy only. These calculations typically yield higher values
of \chisq/\ndf\ of the fits and the scale choice has a higher impact on
the NLO results. These observations emphasize the improved
perturbative convergence of the NNLO calculations as compared to NLO
accuracy.

\begin{figure}[t!b!hp]
\begin{minipage}{0.48\linewidth}
\includegraphics[width=0.90\textwidth]{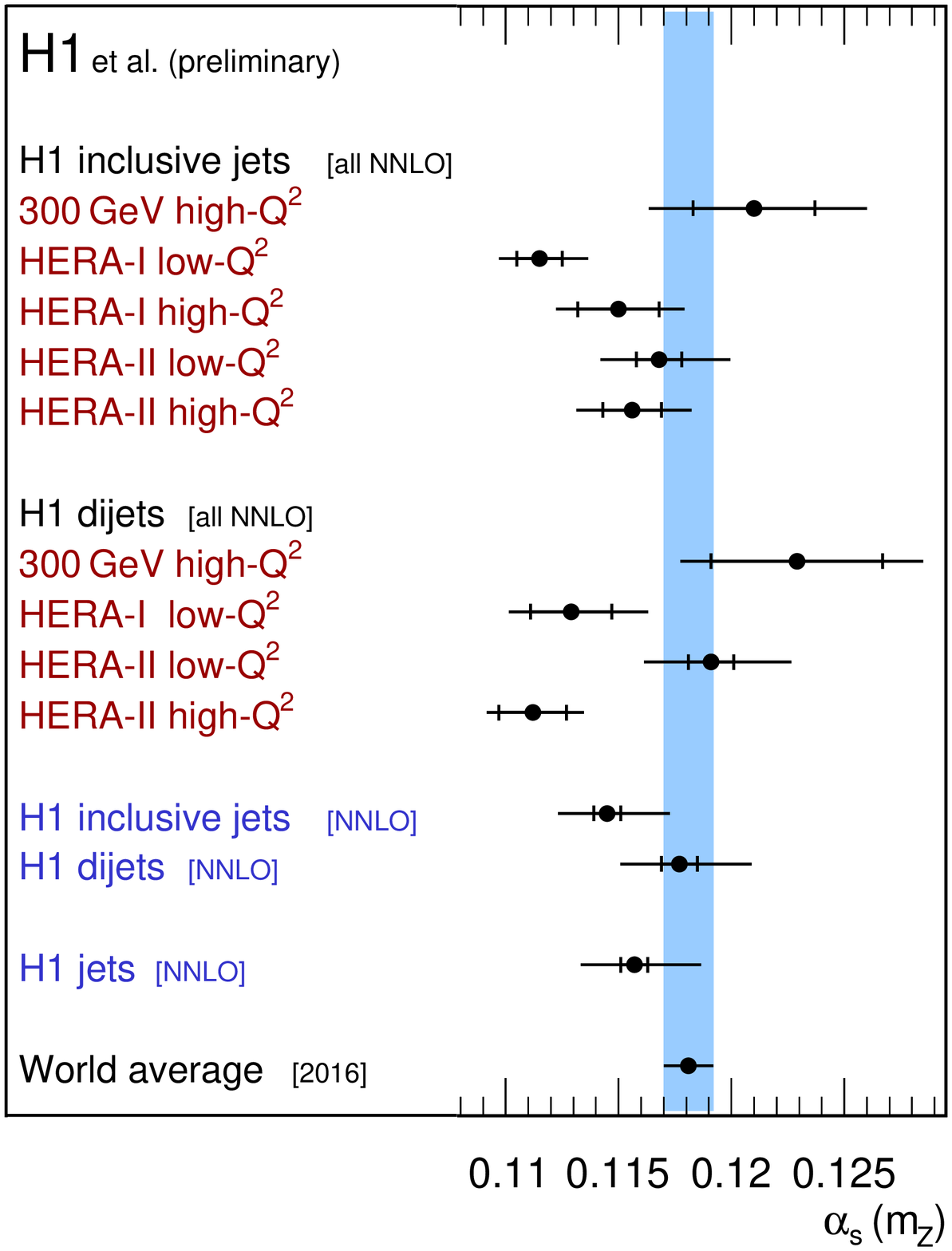}
\caption{   
 Summary of the values of \asmz\ obtained from fits to the individual
 data sets and from fits to multiple data sets. The inner errors bars
 indicate the experimental uncertainty and the outer error bars the total uncertainty.
}
\label{fig:plot2a}
\end{minipage}
\hskip0.03\linewidth
\begin{minipage}{0.48\linewidth}
\begin{center}
   \includegraphics[width=0.90\textwidth]{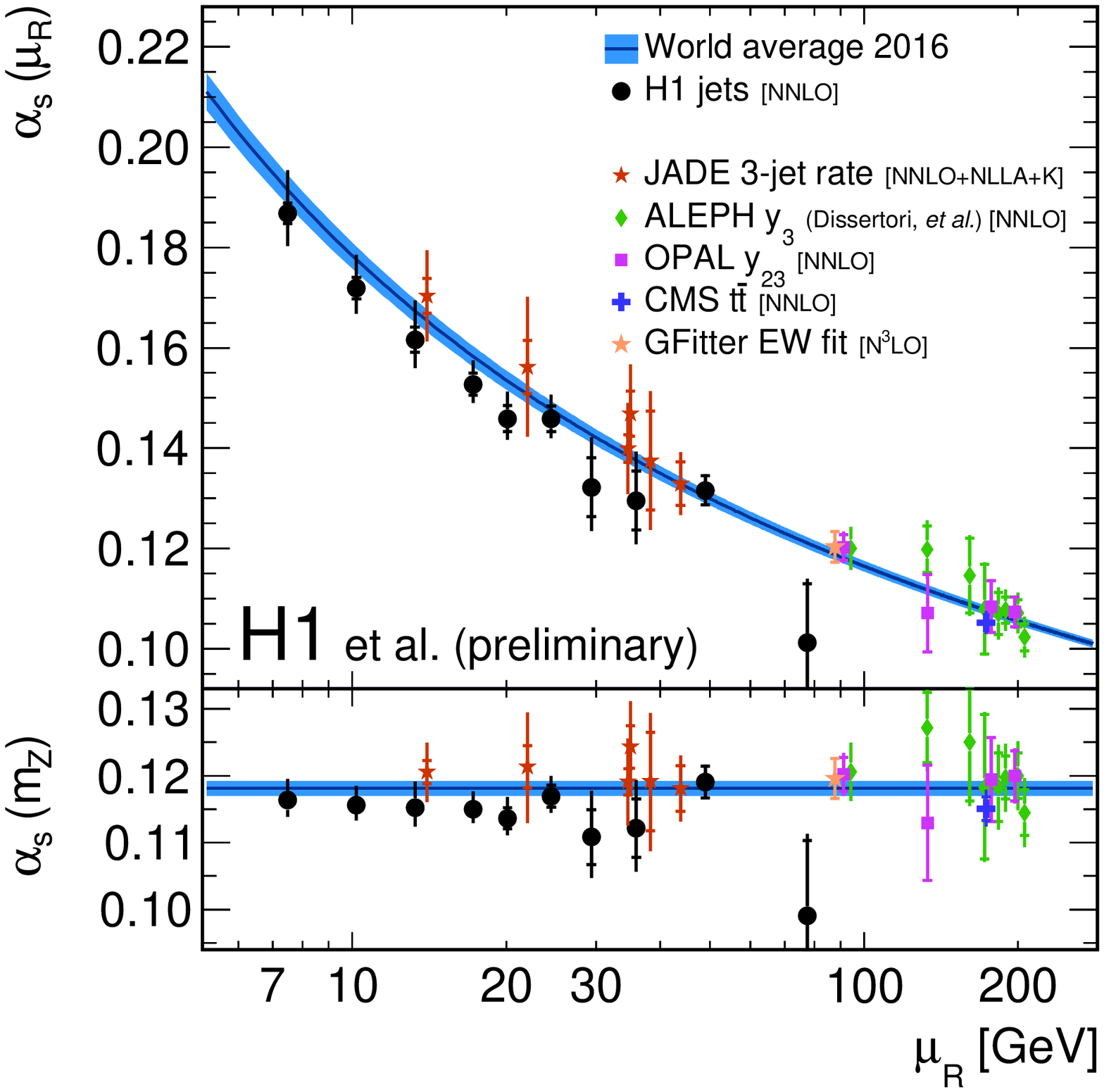}
\caption{
  Values of \asmz\ obtained from fits to `H1 jets' data points with similar values of
  \mur\ (full circles) in comparison to values from other experiments
  and processes, where all values are obtained at least in NNLO accuracy.
  The fitted values of \asmz\ are translated to \asmur\ using the solution of
  the QCD renormalisation group equation as they also enter the calculations.
  The inner error bars display the experimental uncertainties and the
  outer error bars indicate the total  uncertainties.
}
\label{fig:plot2b}
\end{center}
\end{minipage}
\end{figure}
The values for \asmz\ obtained from fits to the individual data sets
are displayed in figure~\ref{fig:plot2a} and compared to the world
average value of $\asmz=0.1181\pm0.0011$~\cite{Olive:2016xmw}.
The results obtained when using only inclusive jet data or only
dijet data are also shown. An overall reasonable
consistency between the results from the individual data sets is
found.

A fit to all H1 jet cross section data (denoted `H1 jets'), where
however the HERA-I dijet cross sections are excluded from the fit because
their statistical correlations to the inclusive jets are not precisely
known, yields a value of $\chisq/\ndf = 1.03$ for 203 data points and
the value of the strong coupling constant \asmz\ is determined to
\begin{equation}
  \asmz = 0.1157\,(6)_{\rm exp}\, (3)_{\rm had}\, (6)_{\rm PDF}\, (12)_{\rm PDF\as}\, (2)_{\rm PDFset}\, (^{+27}_{-21})_{\rm scale}~.
  \nonumber
\end{equation}
This is consistent with the world average and with fits of the
individual data sets. 

The running of the strong coupling constant as a function of
the renormalisation scale \mur, is studied by repeating the fit for
groups of data points with comparable values of \mur.
The resulting values of \asmz\ and \asmur\ are displayed at a representative value
\mur\ for the given range in figure~\ref{fig:plot2b}. The results confirm the
expectations from the QCD renormalisation group equation within the
accessible range in \mur\ of approximately 7 to 90\,\GeV. The \as-values
are also compared
to \as-determinations at NNLO in other reactions at similar scales and consistency is found.

\section{Summary and conclusion}
The strong coupling constant is determined in a fit of new
next-to-next-to-leading order (NNLO) QCD predictions to inclusive
jet and dijet cross section measurements by the H1 experiment as
$\asmz=0.1157\,(6)_{\rm exp}\,(^{+31}_{-26})_{\rm theo}$~\cite{thisprelim}, which is
in consistency with the world average value.
The running of the strong coupling constant is probed over one order
of magnitude and consistency with the QCD expectation is found.
The NNLO calculations reduce significantly  the dominating
theoretical uncertainty in comparison to previously available NLO
calculations. 
The experimental uncertainty is reduced by
considering the entire set of inclusive jet and dijet cross section measurements by the H1 experiment.



\begingroup

\endgroup

\end{document}